\documentclass[reprint,
superscriptaddress,
 amsmath,amssymb,
 aps,
 apl,
]{revtex4-1}

\usepackage{graphicx}
\usepackage{dcolumn}
\usepackage{bm}
\usepackage{xcolor} 
\usepackage{filecontents}
\usepackage{multirow} 
\usepackage{xcolor}
\RequirePackage[T1]{fontenc}
\RequirePackage[utf8]{inputenc}
\RequirePackage{lmodern}
\usepackage{hyperref}

\begin{document}

\title{Improved placement precision of implanted donor spin qubits in silicon\\using molecule ions
}

\author{Danielle Holmes}
\email[Corresponding author: ]{danni.holmes@unsw.edu.au}
\affiliation{Centre for Quantum Computing and Communication Technology, School of Electrical Engineering and Telecommunications, UNSW
Sydney, NSW 2052, Australia}
\author{Benjamin Wilhelm}
\affiliation{Centre for Quantum Computing and Communication Technology, School of Electrical Engineering and Telecommunications, UNSW
Sydney, NSW 2052, Australia}
\author{Alexander M. Jakob}
\affiliation{Centre for Quantum Computing and Communication Technology, School of Physics, The University of Melbourne,
Melbourne, VIC 3010, Australia}
\author{Xi Yu}
\affiliation{Centre for Quantum Computing and Communication Technology, School of Electrical Engineering and Telecommunications, UNSW
Sydney, NSW 2052, Australia}
\author{Fay E. Hudson}
\affiliation{Centre for Quantum Computing and Communication Technology, School of Electrical Engineering and Telecommunications, UNSW
Sydney, NSW 2052, Australia}
\affiliation{Diraq, Sydney, NSW 2052, Australia}
\author{Kohei M. Itoh}
\affiliation{School of Fundamental Science and Technology, Keio University, Minato City, Yokohama, Japan}
\author{Andrew S. Dzurak}
\affiliation{Centre for Quantum Computing and Communication Technology, School of Electrical Engineering and Telecommunications, UNSW
Sydney, NSW 2052, Australia}
\affiliation{Diraq, Sydney, NSW 2052, Australia}
\author{David N. Jamieson}
\affiliation{Centre for Quantum Computing and Communication Technology, School of Physics, The University of Melbourne,
Melbourne, VIC 3010, Australia}
\author{Andrea Morello}
\affiliation{Centre for Quantum Computing and Communication Technology, School of Electrical Engineering and Telecommunications, UNSW
Sydney, NSW 2052, Australia}

\begin{abstract}
Donor and dot spins in silicon-28 ($^{28}$Si) are among the most performant qubits in the solid state, offering record coherence times and gate fidelities above 99\%. Donor spin qubits can be fabricated using the semiconductor-industry compatible method of deterministic ion implantation.
Here we show that the precision of this fabrication method can be boosted by implanting molecule ions instead of single atoms. The bystander ions, co-implanted with the dopant of interest, carry additional kinetic energy and thus increase the detection confidence of deterministic donor implantation employing on-chip single ion detectors to signal the electron-hole pairs induced by ion implants. This allows the placement uncertainty of donor qubits to be minimised without compromising on detection confidence. We investigate the suitability of phosphorus difluoride (PF$_2^+$) molecule ions to produce high quality P donor qubits. Since $^{19}$F nuclei have a spin of $I=1/2$, it is imperative to ensure that they do not hyperfine couple to P donor electrons as they would cause decoherence by adding magnetic noise. Using secondary ion mass spectrometry, we confirm that F diffuses away from the active region of qubit devices while the P donors remain close to their original location during a donor activation anneal. PF$_2$-implanted qubit devices were then fabricated and pulsed electron spin resonance (ESR) measurements were performed on the P donor electron. A pure dephasing time of $T_2^*= 20.5\pm0.5$ $\mu$s and a coherence time of $T_2^{\text{Hahn}}=424\pm5$ $\mu$s were extracted for the P donor electron- values comparable to those found in previous P-implanted qubit devices. Closer investigation of the P donor ESR spectrum revealed that no $^{19}$F nuclear spins were found in the vicinity of the P donor. Molecule ions therefore show great promise for producing high-precision deterministically-implanted arrays of long-lived donor spin qubits.
\end{abstract}
                           
\maketitle

\section{Introduction}

Semiconductor spin qubits \cite{chatterjee2021semiconductor,burkard2023semiconductor} are  appealing platforms for the construction of scalable quantum computers. On the classical engineering side, they offer compatibility with standard semiconductor microelectronics manufacturing processes; on the quantum side they have recently crossed the threshold of one- and two-qubit logic gate fidelities exceeding 99\% \cite{mkadzik2022precision,noiri2022fast,xue2022quantum,mills2022two}. Among the possible physical qubit implementations in semiconductors, donor (also referred to as `dopant' or `impurity') atoms in silicon were the first to be proposed \cite{kane1998silicon} and experimentally demonstrated \cite{pla2012single,pla2013high} as spin qubits. With the introduction of isotopically enriched $^{28}$Si host crystals, which possess no nuclear spins apart from a small fraction (typically $<0.1\%$ \cite{itoh2014isotope}) of residual $^{29}$Si atoms, donor spin qubits have achieved extraordinary values of spin coherence times, both in bulk \cite{tyryshkin2012electron,saeedi2013room} and in single-atom nanoscale devices \cite{muhonen2014storing}. As eloquently stated in a recent review \cite{stano2022review}, `Impurity spins hold record coherence times in each category where data for them exist'. 

The key challenge for donor spin qubits is the development of fabrication methods to place and control individual atoms with high precision, within scalable and manufacturable nanostructures. There exist two different ways to do so. One involves the use of scanning tunneling microscope (STM) lithography \cite{fuechsle2012single,wyrick2022enhanced} to define a hydrogen mask through which phosphine (PH$_3$) molecules are absorbed onto Si, and the P atoms are subsequently incorporated in the crystal. This method yields sub-nanometer placement precision, and a very clean crystalline environment for the spins \cite{kranz2020exploiting}, but typically results in multiple P atoms incorporated in tightly-spaced clusters \cite{buch2013spin}. 
The other, adapted from the classical semiconductor industry \cite{poate2002ion}, involves the implantation of individual dopant atoms into the crystal \cite{van2015single,jamieson2017deterministic}. This method retains compatibility with the standard metal-oxide-semiconductor (MOS) fabrication workflow, and has been developed to fabricate arrays of single atoms, implanted and detected with $\approx 99.9\%$ confidence \cite{jakob2022deterministic}. All demonstrations of coherent single-atom spin control to date have been achieved with ion-implanted donors \cite{morello2020donor}. However, ion straggling causes $>10$~nm uncertainties in the exact location of the implanted atoms, which calls for creative ways to design two-qubit logic gates that are insensitive to such uncertainty \cite{kalra2014robust,madzik2021conditional,tosi2017silicon}. In this work, we demonstrate a technique that reduces the straggle-induced uncertainty in the placement of implanted donors and enables the production of donor qubits closer to the Si surface for stronger coupling to a single electron transistor (SET) for fast readout and initialisation \cite{morello2010single,mohiyaddin2013noninvasive}, without compromising on the detection confidence. 

In order to produce the large-scale arrays of implanted donor spin qubits required for quantum error correction architectures to enable fault-tolerant quantum computers, a high detection confidence is essential to ensure the accurate counting of donor ions into the Si substrate. This requirement poses a challenge, since a low ion implantation energy ($\sim$8-35 keV) is required to place donor qubits at shallow depths ($\sim$7-20 nm below the gate oxide interface) to allow for spin readout via tunneling to an SET and electrostatic tuning by surface nanoelectronics \cite{mohiyaddin2013noninvasive,jakob2022deterministic}. Low implantation energies reduce the detection signal, which consists of the charge induced by the generation of electron-hole pairs by electronic stopping processes of the ion as it slows down in the Si substrate. The detection confidence of a single implanted ion is determined by the signal to noise ratio of the ion beam induced charge (IBIC) signal. Recent advances in single ion detector design, materials quality and low-noise electronics have led to the demonstration of detection fidelities up to 99.85 \% for single implanted P donors at an energy of 14 keV \cite{jakob2022deterministic}. However, the placement of donors closer to the surface and the reduction of straggle requires a lower implant energy.

\begin{figure}[h]
\centering
\includegraphics[scale=0.58]{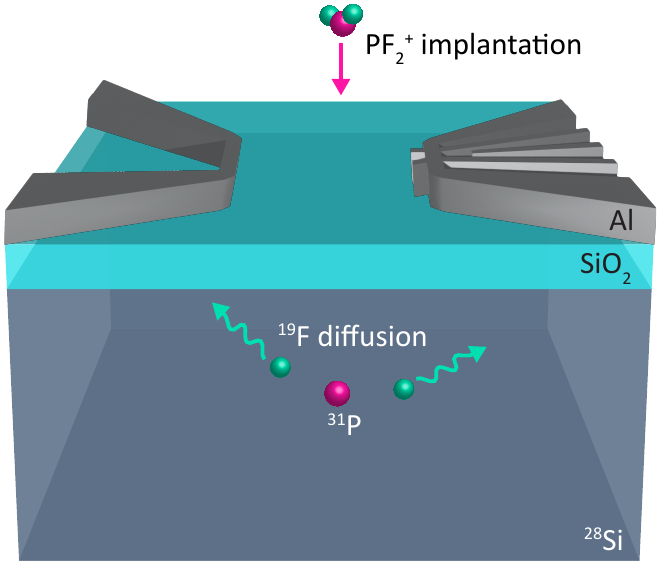}
\caption{A donor spin qubit device fabricated using the implantation of PF$_2^+$ molecule ions. The bystander $^{19}$F ions diffuse away from the active region of the device upon a donor activation anneal, while the $^{31}$P donor qubit remains close to the original stopping location.}
\label{fig:PF2device}
\end{figure}

Here we demonstrate the use of molecule ions as a way to retain large implantation energies, and thus IBIC signals, while reducing the implantation depth and straggle of the donor qubit. Molecule ions are commonly used in the microelectronics industry for the formation of ultra-shallow junctions, crucial in the fabrication of metal-oxide-semiconductor field-effect transistors with decreasing device dimensions for increased component density. For example, the boron difluoride (BF$_2$) molecule ion is favoured over atomic boron (B) ions to produce shallow implanted p-type regions in Si due to its smaller projected range, its ability to induce amorphization of Si to suppress ion channeling \cite{wilson1983boron} and to eliminate damage during solid-phase epitaxial regrowth \cite{mayer1970ion,holmes2021isotopic} and the ability of the co-implanted F to reduce boron transient enhanced diffusion \cite{downey1998effect}. In this work, we employ phosphorus difluoride (PF$_2^+$) molecule ions to fabricate P donor qubits, shown schematically in Fig. \ref{fig:PF2device}. The P donor carries a fraction of the molecule ion's energy in proportion to its mass. Upon impact, the molecule ion breaks apart since the implant energy is much larger than the molecular binding energy, effectively co-implanting separate atomic ions at the same exact impact site. P donors can be placed to a depth of $\sim$15 nm below an 8 nm gate oxide, suitable for spin initialisation and readout, by implanting PF$_2^+$ molecule ions at an energy of 19.5 keV. Under these conditions, the P donor carries an energy of $\sim$9 keV, resulting in a low straggle of $<8$ nm. The F bystander ions within the molecule ion carry the remaining energy, and boost the IBIC signal further above the noise threshold of the detector.

Fluorine nuclei have a spin ($I=1/2$) and a large gyromagnetic ratio. Spin fluctuations of $^{19}$F nuclei close to the P donor qubit (i.e. at distances of order the Bohr radius, 1.22~nm for P in Si \cite{jamieson2017deterministic}) would cause perturbations in the local magnetic field at the donor site, resulting in a time-varying qubit resonance frequency and consequent reduction of the electron spin coherence time \cite{witzel2010electron}. Fortunately, F is known to be a fast diffuser in Si \cite{jeng1992anomalous}. We show that F diffuses away from the active region of qubit devices while the P donors remain close to their original location during a donor activation anneal, using secondary ion mass spectrometry (SIMS). 

For the first time, we demonstrate the operation of a donor qubit produced using the implantation of a molecule ion. The suitability of PF$_2$ for making long-lived P donor qubits was confirmed by fabricating PF$_2$ implanted qubit devices and performing electron spin resonance (ESR) measurements. The P donor electron was not found to couple to any $^{19}$F nuclear spins.

\section{Experiment}

Intrinsic crystalline Si substrates were implanted with PF$_2^+$ molecule ions with various implantation energies and fluences using a Colutron ion implanter with a gas source consisting of 5\% PF$_5$ diluted in 95\% Ar, as summarised in Table \ref{table:Implantparameters}. Samples were implanted at room temperature, with a beam line  pressure of $\sim$$5\times10^{-8}$ Torr, with a substrate tilt of 7$^\circ$ relative to the incident beam axis to suppress ion channeling.

{\renewcommand{\arraystretch}{1.3}
\begin{table}[h]
\centering
 \begin{tabular}{|c|c|c|c|} 
 \hline
  & Sample A & Sample B & Sample C \\
 \hline
Implant species &  PF$_2^+$ &   PF$_2^+$    & PF$_2^+$ \\
 \hline
 Implant energy & \multirow{2}{*}{20} & \multirow{2}{*}{20}  & \multirow{2}{*}{19.5} \\
 (keV) &  &  & \\
 \hline
 Implant fluence & \multirow{2}{*}{$3\times10^{14}$} & \multirow{2}{*}{$3\times10^{14}$} & \multirow{2}{*}{$2\times10^{11}$} \\
 (cm$^{-2}$) & & &\\
 \hline
  Anneal & \multirow{2}{*}{None} & \multirow{2}{*}{1050, 5, Ar} & \multirow{2}{*}{1000, 5, N$_2$}\\
 ($^\circ$C, s, ambient) & & &\\
 \hline
   Measurement & SIMS & SIMS & ESR\\
 \hline
\end{tabular}
\caption[PF$_2$ implant parameters]{Implant parameters for the samples measured in this work.}
\label{table:Implantparameters}
\end{table}}

To determine the diffusion of P and F in Si due to a rapid thermal anneal (RTA), samples A and B were measured using SIMS. To produce samples A and B, intrinsic natural $<$100$>$ Si substrates with a native oxide surface layer were implanted with 20 keV PF$_2^+$ ions at a fluence of 3$\times10^{14}$ cm$^{-2}$. This implantation results in a peak concentration of P and F of $1.5\times10^{20}$ cm$^{-3}$ and $3\times10^{20}$ cm$^{-3}$, respectively, at a depth of $\sim$15 nm below the surface. The implant fluence was chosen to ensure the P and F concentrations are well above the minimum detectable limit for SIMS analysis and is around 3 orders of magnitude higher than those employed for qubit fabrication. Sample A was left as-implanted, while sample B was annealed at 1050 $^\circ$C for 5 s in an argon atmosphere. Due to the shallow implant profile, both samples were capped with a 50 nm Pt layer to increase the accuracy of the SIMS analysis near the Si surface. The details of the SIMS measurement is provided in the following.

To determine the suitability of PF$_2$ molecule ions for producing long-lived P donor qubits, sample C had a nanoelectronic qubit device fabricated on-chip to perform ESR of the P donor qubit. To produce sample C, a $^{28}$Si epilayer with a high quality 8 nm gate oxide was implanted with 19.5 keV PF$_2^+$ ions at a fluence of $2\times10^{11}$ cm$^{-2}$ through a 100 $\mu$m diameter aperture into a 90 nm $\times$ 100 nm implant window. This implantation results in a peak concentration of P and F of $\sim$$1.1\times10^{17}$ cm$^{-3}$ and $\sim$$1.9\times10^{17}$ cm$^{-3}$ and an average implanted depth of $\sim$8 nm and $\sim$7 nm below the Si/SiO$_2$ interface, respectively. This sample was then given an RTA at 1000 $^\circ$C for 5 s in a nitrogen atmosphere. Surface nanoelectronics were then fabricated as standard for our qubit devices using multiple layers of electron beam lithography and aluminium deposition. This device was then packaged and cooled to around 19 mK in a dilution refrigerator. Details of the ESR measurements are provided in the following.

\section{Minimising qubit placement uncertainty}

In other work, we found that 9 keV P$^+$ ions implanted into on-chip single ion detectors resulted in a detection confidence of $\sim$98.6\%. This implantation energy can be simulated using the Stopping and Range of Ions in Matter (SRIM) Monte-Carlo model \cite{ziegler2010srim} to have an average range of 16 nm below the sample surface and a longitudinal straggle of $\sim$8 nm. If we instead utilise PF$_2^+$ ions, implanted at an energy of 19.5 keV to maintain the fraction of energy carried by P to be $\sim$9 keV, we could boost the detection confidence to over 99.95\%. This result clearly shows the ability of the bystander F ions to shift the IBIC signal further above the noise floor of the detector without increasing the average implanted range or straggle of the P donor.

\begin{figure}[h]
\centering
\includegraphics[scale=0.58]{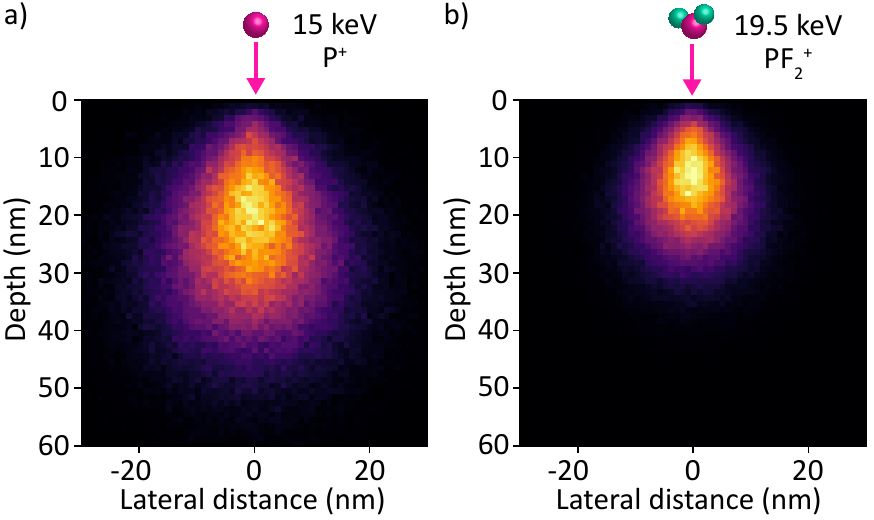}
\caption{SRIM simulation of the P ion distribution resulting from the implantation of a) 15 keV P$^+$ ions and b) 19.5 keV PF$_2^+$ ions into Si. The colour scale represents the probability distribution of the P donors. Effects from ion channelling were neglected.}
\label{fig:SRIM}
\end{figure}

Without the use of molecule ions, to achieve a detection confidence of $\sim$99.95\%, we would require a P implantation energy of $\sim$15 keV. This increased energy would result in an increased average implanted range of $\sim$24 nm and an increased longitudinal straggle of $\sim$11.5 nm, with a distribution shown in Fig \ref{fig:SRIM}a simulated by implanting 200,000 P ions in Si at an energy of 15 keV using SRIM. The corresponding simulated P distribution that results from the implantation of 19.5 keV PF$_2$ molecule ions is shown in Fig. \ref{fig:SRIM}b. These SRIM simulations visualise the ability of PF$_2$ molecule ions to minimise the donor qubit implantation depth and placement uncertainty without compromising on the single ion detection confidence. This has positive implications on the scalability of deterministically implanted arrays of donor spin qubits in Si, engineered with high placement precision to achieve the desired coupling strengths of donor electrons both to their neighbouring qubits and to surface electronics for spin readout and initialisation. 

\section{Fast diffusion of fluorine}

As standard practice after the implantation of P donor qubits, samples are given a rapid thermal anneal at 1000 $^\circ$C for 5 s in a nitrogen atmosphere. This is to repair the damage to the Si crystal caused by ion implantation and to ensure that the P donor is substitutional in the Si lattice and is electrically active \cite{sedgwick1983short,saito1998ultra}. The diffusion of both P and F dopants in Si as a result of this rapid thermal anneal was determined by measuring the concentration depth profiles with SIMS, before and after annealing. 

\begin{figure}[h]
\centering
\includegraphics[scale=0.75]{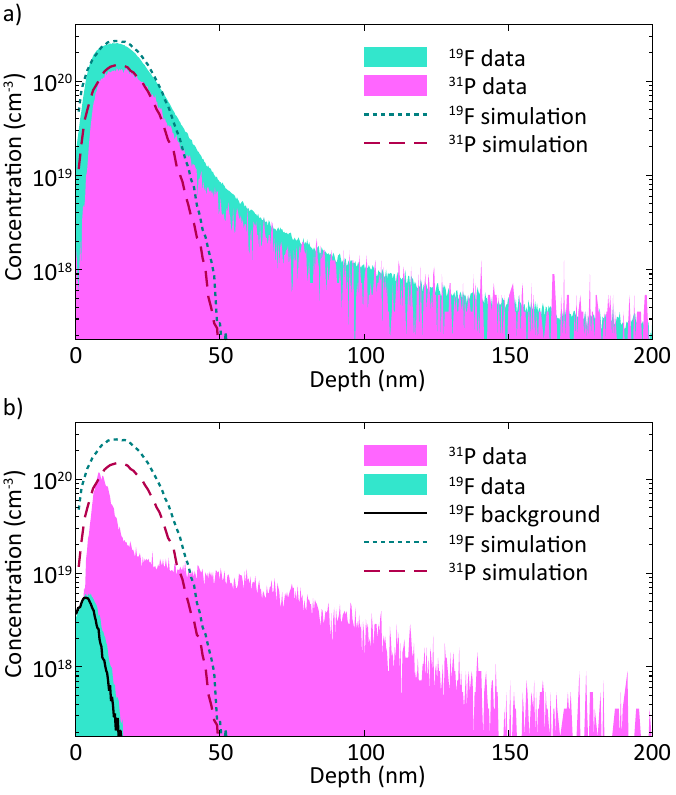}
\caption{SIMS measurements showing the concentration of P and F as a function of depth below the Si surface for a) sample A and b) sample B. The coloured dashed lines show the depth profiles of P and F simulated using SRIM for an implantation of 20 keV PF$_2$ at a fluence of $3\times10^{14}$ cm$^{-2}$. The black dashed line in b) shows the background F concentration present in sample B outside of the implanted region.}
\label{fig:SIMS}
\end{figure}

Samples A and B were analysed with time-of-flight secondary ion mass spectrometry (TOF-SIMS) in negative polarity by Bi$^+$ ions at 30 keV, sputtered by Cs$^+$ ions at 500 eV. The concentration of both P and F as a function of depth below the Si surface, before and after the rapid thermal anneal required for donor activation, is given in Fig. \ref{fig:SIMS}.

The implantation profiles of both F and P before a thermal anneal are given in Fig. \ref{fig:SIMS}a. The measured concentration profiles are in good agreement with the simulated profiles produced using SRIM (dashed lines) in the near-surface region up to a depth of $\sim$30 nm. The simulated concentration of F and P is lower than measured at greater depths due to the SRIM simulation not taking into account ion channelling into the crystalline Si target. 

A comparison of the SIMS data for P shown in Fig. \ref{fig:SIMS}a and Fig. \ref{fig:SIMS}b shows that after the RTA, the majority of P remains in the near surface region (with a peak concentration $\sim$9 nm below the Si surface). A significant fraction of P donors are diffused towards greater depths below the Si surface (by up to $\sim$100 nm) as a result of the donor activation anneal. This diffusion is likely enhanced by the elevated presence of point defects introduced in Si due to the damage caused by ion implantation \cite{oehrlein1984diffusion}. This transient enhanced diffusion is likely to be less significant in donor qubit devices, where the donor implantation fluence is around 3 orders of magnitude lower \cite{cowern1986transient}. The total integrated area under the $^{31}$P concentration depth profile for the annealed sample gives an equivalent fluence of $\sim$$1.6\times10^{14}$ cm$^{-2}$, just over half the implanted fluence in the non-annealed sample. This reduced amount of P may be due to the resolution of the SIMS measurement not being high enough to accurately measure the spike in the P concentration that has diffused to the surface. Additionally, a fraction of the P that diffused to the Si surface may have evaporated through the native silicon oxide layer during the high temperature RTA. This outgassing effect is likely to be less significant for qubit devices which have a high quality 8 nm gate oxide on the surface. These measurements show that a significant concentration of P remains in the active region of a donor qubit device- at a depth at which donor electrons can tunnel couple to a single electron transistor island to enable spin initialisation and readout.

On the contrary, a comparison of the SIMS data for F shown in Fig. \ref{fig:SIMS}a and Fig. \ref{fig:SIMS}b shows that F atoms diffuse significantly away from their as-implanted profile during the RTA. After annealing, the concentration of F found in the active region of a donor qubit device has reduced all the way down to the background level found in the Si substrate. A concentration of F up to $6\times10^{18}$ cm$^{-3}$ is measured at $\sim$0-15 nm below the Si surface after the RTA. However, this corresponds almost identically to the depth profile of the background F measured in a region of the same sample that was not implanted with PF$_2$ ions (black dashed line in Fig. \ref{fig:SIMS}b). This measured near-surface F is therefore not caused by the implantation of PF$_2^+$ ions but is instead likely due to forward recoils of F, present as contamination in the sputter-deposited Pt capping layer or the Si surface, during the SIMS measurement. F implanted at low energies and fluences has been shown to diffuse towards the surface of Si where it outgasses, reducing the amount of F present in the sample \cite{jeng1992anomalous}. In qubit devices with a high quality gate oxide, a fraction of the implanted F may be gettered at the Si/SiO$_2$ interface. F has been shown to act like hydrogen in a forming gas anneal by passivating dangling bonds at the interface through the formation of Si-O-F and Si-F complexes, which would have the beneficial effect of reducing noise in the qubit device \cite{ono1993segregation}.

The SIMS data shows that after the RTA required to activate the implanted P donors, there is no additional concentration of F in the active region of a qubit device due to the implantation of PF$_2^+$ molecule ions. This result shows the PF$_2$ molecule ion to be a promising alternative to introducing long-lived P donor qubits into a Si device via implantation as they should not contribute any significant levels of spin-containing nuclei into the qubit environment.

\section{P donor spin qubit measurements}

To determine the suitability of PF$_2$ molecule ions for producing high quality implanted P donor spin qubits, a nanoelectronic qubit device \cite{mkadzik2022precision} was fabricated on the surface of sample C and one of the P donors introduced via implantation of PF$_2^+$ was operated as an electron spin qubit \cite{pla2012single}. Sample C was mounted on the mixing chamber of a dilution refrigerator, with a base temperature $\sim$19~mK and an electron temperature $\sim$150~mK. An external magnetic field with strength $B_0\approx1.4$ T was applied along the $z$-axis using a superconducting solenoid, resulting in a Zeeman splitting of the electronic and nuclear spin states. The Hamiltonian of the donor system can then be written as \cite{pla2012single,pla2013high}

\begin{equation}
    H = \gamma_e B_0 \hat{S}_z - \gamma_n^{\text{P}} B_0 \hat{I}_{z}^{\text{P}} +\sum_{k}A^k \hat{\boldsymbol{S}}.\hat{\boldsymbol{I}}^k ,
\end{equation}
where the donor electron and $^{31}$P nuclear spins have gyromagnetic ratios $\gamma_e \approx 27.97$ GHz/T \cite{feher1959electron} and $\gamma_n^{\text{P}} \approx 17.23$ MHz/T \cite{steger2011optically}, respectively, and $\hat{S}_z$ and $\hat{I}_{z}^{\text{P}}$ are the electron and $^{31}$P nuclear spin projection operators, respectively. The donor electron is hyperfine coupled to $k$ nuclei residing within its Bohr radius, each with a coupling strength $A^k$ and a nuclear spin $\hat{\boldsymbol{I}}^k$. This final term is often simplified to contain solely the hyperfine coupling between the donor electron and the $^{31}$P nucleus, $A^{\text{P}} \hat{\boldsymbol{S}}.\hat{\boldsymbol{I}}^{\text{P}}$. The hyperfine couplings to other nearby nuclei, such as residual $^{29}$Si, are orders of magnitude smaller \cite{ivey1975ground} and their effect is usually handled as a source of decoherence \cite{witzel2010electron}.

Nanoelectronic structures on the surface of the chip provide electrostatic control of the donors, create a single-electron transistor (SET) charge sensor, and deliver microwave and radio-frequency signals through a broadband antenna (Fig. \ref{fig:PF2device}). With this set-up, we can perform single-shot electron spin readout \cite{morello2010single}, and high fidelity (approximately 99.9\%) single-shot quantum nondemolition readout of the nuclear spins \cite{pla2013high}, as well as nuclear magnetic resonance (NMR) and ESR \cite{pla2012single} on all spins involved.

\begin{figure}[h]
\centering
\includegraphics[scale=0.7]{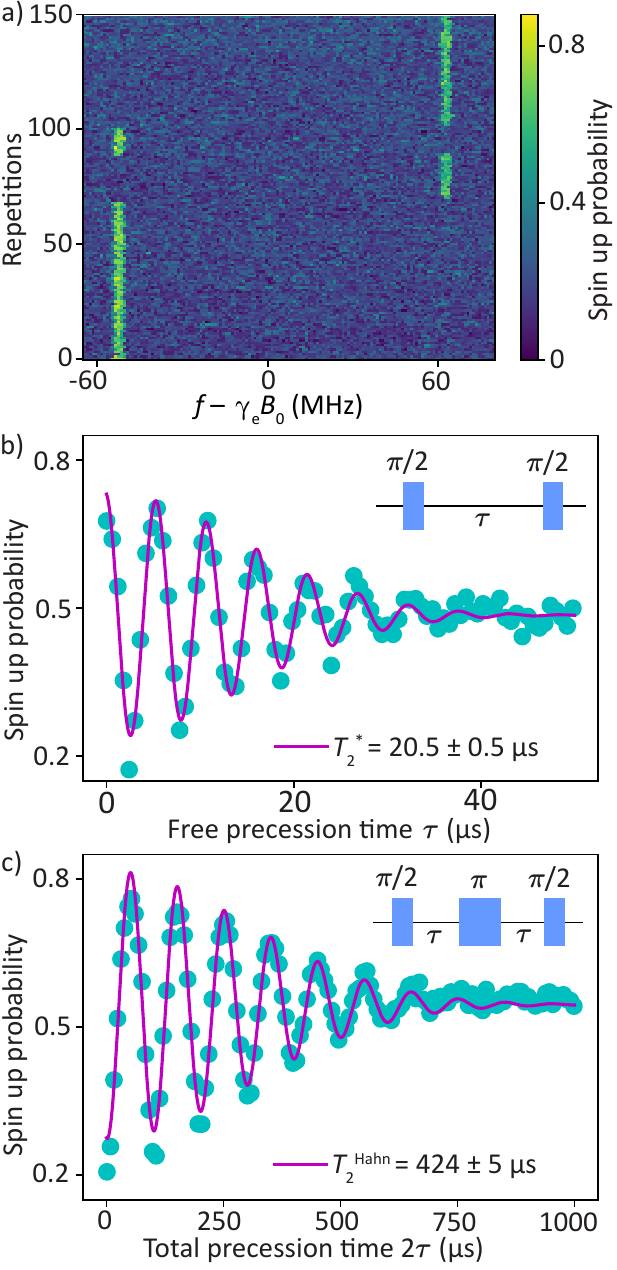}
\caption{a) Adiabatic ESR spectrum of a $^{31}$P donor electron in Si. Two ESR peaks are visible with a splitting of $\sim$115 MHz due to the hyperfine coupling to the $I=1/2$ nuclear spin of the $^{31}$P donor. b) Ramsey measurement on the $^{31}$P donor electron with the corresponding pulse sequence shown in the top right. The fit yields a pure dephasing time $T_2^*=20.5\pm0.5$ $\mu$s. c) Hahn echo measurement on the $^{31}$P donor electron with the corresponding pulse sequence shown in the top right. The data is fit with a sinusoidal Gaussian decay, yielding a coherence time $T_2^{\text{Hahn}}= 424 \pm 5$ $\mu$s.}
\label{fig:aESR}
\end{figure}

The presence of a $^{31}$P donor in our qubit device was first confirmed by acquiring an adiabatic ESR spectrum, as shown in Fig. \ref{fig:aESR}a. The spectrum was obtained by sweeping the centre frequency, $f$, of an applied microwave chirp signal designed to adiabatically invert the electron spin \cite{laucht2014high}, around $\gamma_e B_0 \approx 39$ GHz and measuring the probability of the electron being found in the spin up state. A high spin up proportion (`ESR peak') is measured when the frequency range covered by the adiabatic sweep encompasses the ESR frequency. The two observed ESR peaks are split by the hyperfine coupling to an $I=1/2$ $^{31}$P nuclear spin, with a strength of $A^{\text{P}}\approx115$ MHz, which is close to the bulk value for P donors in Si of 117.5 MHz \cite{feher1959electron}. Random switches are observed between the two ESR peaks, corresponding to quantum jumps of the nuclear spin state \cite{pla2013high}. No further splitting of ESR peaks due to hyperfine coupling to additional nuclei was observed on this scale since the adiabatic ESR spectrum shows a linewidth artificially broadended by the width of the frequency sweep, $3$~MHz in this case.

\begin{figure*}[hbt!]
\centering
\includegraphics[scale=0.64]{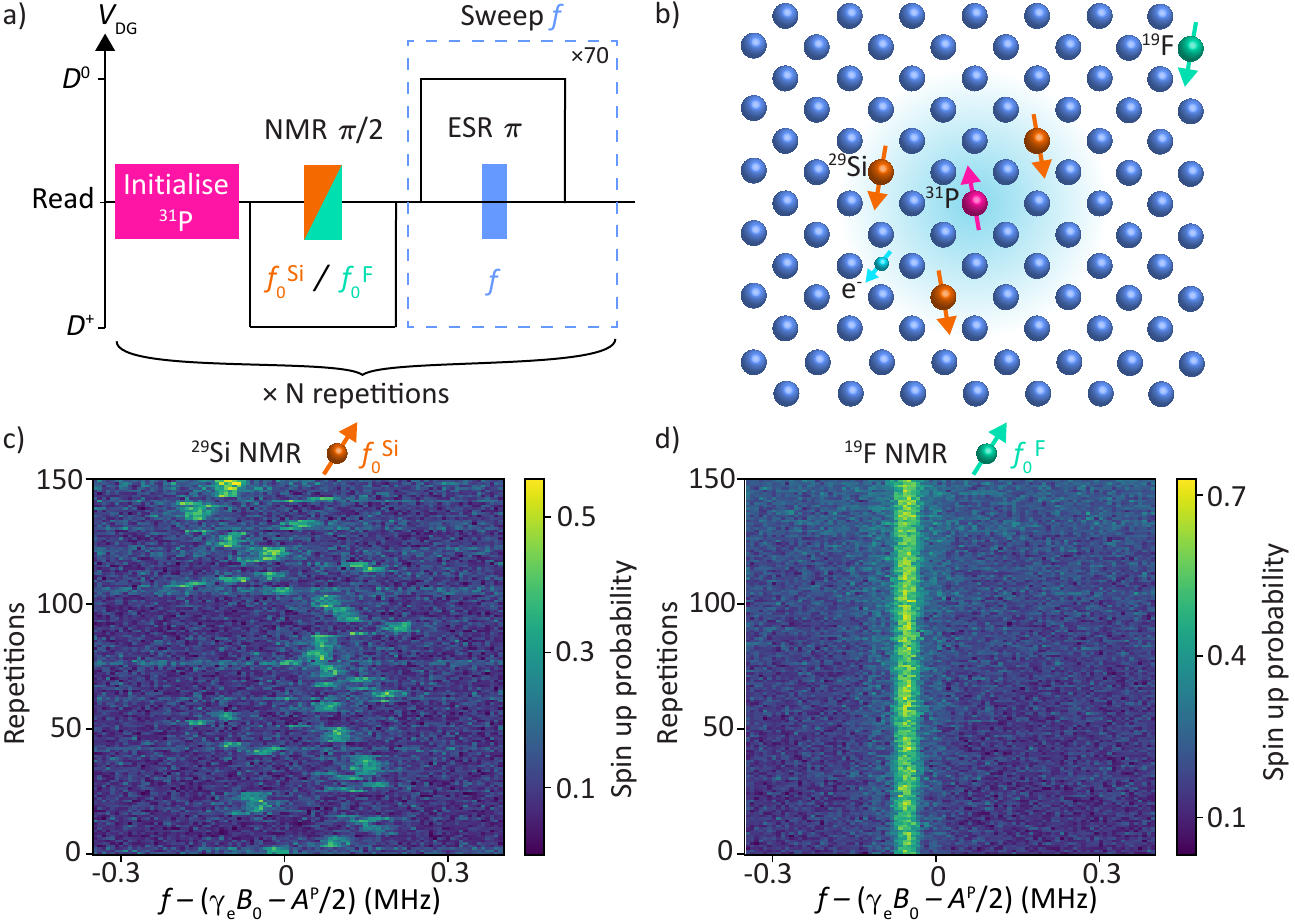}
\caption{a) Pulse sequence used to flip the spins of $^{29}$Si or $^{19}$F nuclei while monitoring the P donor ESR spectrum. b) The Bohr radius of the P donor electron encompasses multiple $^{29}$Si spins but no $^{19}$F spins. c) Coherent ESR spectrum of the lower hyperfine split P donor ESR peak. The P donor electron spin up probability is measured as a function of ESR frequency for multiple repetitions, interleaved with $^{29}$Si NMR $\pi/2$ pulses. The instantaneous ESR frequency is seen to jump frequently between discrete values, determined by the orientation of a few $^{29}$Si nuclear spins, hyperfine coupled to the P electron with strengths of order 100 kHz. d) The same experiment repeated with NMR $\pi/2$ pulses applied at the expected resonance of $^{19}$F nuclei, interleaved with the collection of coherent ESR spectra. The ESR frequency remains completely stable, indicating that no $^{19}$F are coupled to the P donor electron.}
\label{fig:SpinFlips}
\end{figure*}

In order to assess the quality of the PF$_2$ implanted donor spin qubit, the donor electron pure dephasing time, $T_2^*$, and coherence time, $T_2^\text{Hahn}$, were measured using Ramsey and Hahn echo pulse sequences, respectively \cite{pla2012single,muhonen2014storing}. The Ramsey experiment was performed using the pulse sequence given in the inset of \ref{fig:aESR}b. Fitting the data with a sinusoidal Gaussian decay yields an electron pure dephasing time of $T_2^*=20.5\pm0.5$ $\mu$s. The Hahn echo experiment was performed using the pulse sequence given in the inset of \ref{fig:aESR}c. Fitting the data with a sinusoidal Gaussian decay yields an electron coherence time of $T_2^\text{Hahn}=424\pm5$ $\mu$s. These values are comparable to previous donor qubit devices implanted with atomic P ions that have been measured in the same dilution refrigerator. While this result is promising, it does not in itself prove that co-implanted F bystander ions do not negatively impact the P donor qubit since the donor qubit coherence time can depend on the details of the specific device, dilution refrigerator, magnet and measurement setup. 

The pulse sequence to detect weakly-coupled $^{29}$Si or $^{19}$F nuclei is given in Fig. \ref{fig:SpinFlips}a. First, the $^{31}$P donor nucleus is initialised in the spin down state to ensure that the frequency of the lower hyperfine-split ESR peak can be tracked continuously. Next, the donor is prepared in the ionised state, $D^+$, by lowering the voltage on a donor gate, $V_\text{DG}$, to remove the donor electron. An NMR $\pi/2$ pulse is then applied on resonance with $^{29}$Si nuclei or $^{19}$F nuclei, with a frequency of $f_0^{\text{Si}}=\gamma_n^{\text{Si}} B_0$ or $f_0^{\text{F}}=\gamma_n^{\text{F}}B_0$, respectively, to flip approximately half of the coupled nuclei. The donor is then brought into the neutral state, $D^0$, by raising $V_\text{DG}$ to ensure that the donor electron remains bound to the $^{31}$P nucleus. An ESR $\pi$ pulse is then applied at a frequency $f$ and the donor is brought to the read position with $V_\text{DG}$. This is repeated 70 times to measure the spin up probability of the donor electron. The ESR frequency is the swept around the lower hyperfine-split ESR peak at $f=\gamma_e B_0 - A^{\text{P}}/2$. The entire pulse sequence is then repeated $N$ times to track the resonant frequency of the donor electron from the ESR spectra collected after each NMR $\pi/2$ pulse.

First, this measurement was performed using NMR pulses on resonance with $^{29}$Si nuclei, with the data shown in Fig. \ref{fig:SpinFlips}b. The ESR frequency of the P donor electron can be seen to jump between multiple values at short time intervals. The $^{29}$Si NMR pulses that we apply are therefore changing the spin configuration of the $^{29}$Si nuclei that are coupled to the P donor electron. This experiment confirms the presence of $^{29}$Si nuclei within the Bohr radius of the P donor electron.

The measurement was then repeated using NMR pulses on resonance with $^{19}$F nuclei, with the data shown in Fig. \ref{fig:SpinFlips}c. The ESR frequency of the P donor electron can be seen to remain constant over 150 repetitions of the pulse sequence, corresponding to a time period of $\sim$20 minutes. The $^{19}$F NMR pulses that we apply are therefore not changing the spin configuration of any nuclear spins that are coupled to the P donor electron. We note that it is quite normal, in the absence of active NMR stimuli at the $^{29}$Si resonance frequency, for the $^{29}$Si nuclei to remain unchanged over very long time scales, thanks to the phenomenon of `nuclear freezing' \cite{madzik2020controllable}. This investigation leads to the positive result that no $^{19}$F nuclear spins are coupled to the P donor electron spin qubit, to within the intrinsic ESR linewidth of our experiment. This observation confirms the expectation, already supported by the SIMS data (\ref{fig:SIMS}), that the F bystander ions diffuse away from the active region of the qubit device upon performing a donor activation anneal.

\section{Conclusion}

In conclusion, we demonstrate that molecule ions can be employed to boost the placement precision and scalability of deterministic implantation of donor qubits without compromising on qubit quality. The F bystander ions that are co-implanted with P in the PF$_2$ molecule ion boost the detection confidence of implanted single ions by increasing the ion beam induced charge signal further above the detector noise floor. Using secondary ion mass spectrometry, we find that F diffuses away from the active region of qubit devices while the P donors remain close to their original location during a donor activation anneal. We then fabricated PF$_2$ implanted qubit devices and performed ESR measurements to determine that the P donor electron showed coherence times comparable to previous qubits produced using atomic P ion implantation. The ESR spectrum of the P donor electron was then monitored while applying NMR pulses on resonance with particular nuclear species to show that the P donor electron was not coupled to any nearby $^{19}$F nuclear spins, only to residual $^{29}$Si nuclei present in the isotopically enriched $^{28}$Si epilayer. This work demonstrates the suitability of PF$_2$ molecule ions for producing long-lived P donor qubits and supports the employment of molecule ions in the production of scalable qubit arrays.

\begin{acknowledgments}

This research was funded by the Australian Research Council Centre of Excellence for Quantum Computation and Communication Technology (CE170100012) and the US Army Research Office (Contracts no. W911NF-17-1-0200 and W911NF-23-1-0113). We acknowledge support from the Australian National Fabrication Facility (ANFF), the Surface Analysis Laboratory, SSEAU, MWAC, UNSW for SIMS and the support of the International Atomic Energy Agency through the Cooperative Research Program number F11020 ``Ion beam induced spatio-temporal structural evolution of materials: Accelerators for a new technology era''. B.W. and X.Y. acknowledge support from the Sydney Quantum Academy. The views and conclusions contained in this document are those of the authors and should not be interpreted as representing the official policies, either expressed or implied, of the Army Research
Office or the U.S. Government. The U.S. Government is authorized to reproduce and distribute reprints for Government purposes notwithstanding any copyright notation herein.
\end{acknowledgments}

\bibliography{bibliography}

\end{document}